==================

\def\half {{1\over 2}}
\def\be {\begin{equation}}
\def\ee {\end{equation}}
\def\ba {\begin{eqnarray}}
\def\ea {\end{eqnarray}}
\def\ne {\nonumber}

\documentstyle[manuscript,aps]{revtex}
\begin{document}

\title{Maintaining coherence in Quantum Computers.
 }
\author{W. G. Unruh }
\address{
 CIAR Cosmology Program\\
Dept. of Physics\\
University of B. C.\\
Vancouver, Canada V6T 2A6\\}

\maketitle

 ~

 ~

\begin{abstract}
The effect of the inevitable coupling to external degrees of freedom of
a quantum computer are examined. It is found that for quantum calculations
(in which the maintenance of coherence over a large number of states is
important), not only must the coupling be small but the time taken in the
quantum calculation must be less than the thermal time scale, $\hbar/k_B T$.
For longer times the condition on the strength of the coupling to the external
world becomes much more stringent.
\end{abstract}

\section{Introduction}
Quantum computers have recently raised a lot of interest. A number of
papers\cite{1} have argued that quantum computers can solve certain problems
much more efficiently than can classical computers.
  Shor\cite{2} has shown that a quantum computer could solve  the
problem of finding discrete logs (mod N) and of finding the factors of
a
large number N in a time which is a polynomial function of the length
$L$ of the number. For factoring the best known algorithm, the Number Field
Sieve\cite{3} takes a time
of order
$exp(c(L)^{1/3}(\ln(L ) )^{2/3}$, where $c(L)$ is roughly
constant and approximately equal to 2 for large L.
Although this is subexponential, it is worse than any polynomial for large
N.
A crucial feature of
the ability of quantum computers to be more efficient in certain
problems involves
 having the computer be placed in the coherent superposition of
a very large number (exponential in $L$) of ``classical states", and having
the outputs interfere in such
a way that there is a very high probability that on the appropriate reading
of the
output, one would obtain the required answer. One is replacing
exponentiallity in time with exponentiallity in quantum coherence. This
requires that the computer be able to
maintain the coherence during the course of the calculation.
This paper examines this requirement, and examines the constraints
placed on the ability to maintain this coherence in the face of coupling
to external heat baths. Landauer\cite{4} has long emphasized the necessity
of
examining the effect of both imperfections and of the coupling to the
external world of any realistic device
 on the ability of quantum computers to realize their
promise. This paper is thus a first step in that direction.

\section{Decohering Noise}
 I will look at only the simplest model, in which
I ask about the maintenance of coherence in a memory of length $L$. This
does not take into account the effect that the course of the computation
itself would have on the rate of loss of coherence, but I would expect
that only to increase the problem. Thus let us assume that
that the number is represented in the computer as a string of binary digits
of
length of the order of $L=ln(N)$. The memory cells will each be taken
to be two level systems, with each of the two levels having the same energy.
The two states will be take to be the eigenstates of a ``spin" operator
$\sigma_z$.

In a conventional computer, the way in which the calculation is ``kept
on track" is by including dissipation in order to damp out any attempt
by the system to make a transition ( except of course those driven by the
computation)\cite{5}. I will therefore assume that the interaction with
the environment has the two desired eigenstates of the memory as eigenstates
of the interaction.  The environment will be modeled by a massless scalar
field\cite{6} derivatively
coupled to the memory cell, so that the the full Hamiltonian is
\be
H= {1\over 2} \left( (\pi(x)+ \epsilon h(x)\sigma_z)^2+ (\partial_x \phi(x))^2
\right)dx
\ee
(The associated lagrangian has the simple derivative coupling form
\be
L=\half\int \left((\partial_t\phi)^2-(\partial_x\phi)^2 -2\epsilon
h(x)\phi(x)\sigma_z \right) dx.)
\ee
Here $h(x)$ is some interaction range function, and $\pi $ is the momentum
conjugate to $\phi$

The Heisenberg equations of motion are
\ba
\dot \pi= \partial_x^2\phi
\\
\dot \phi = \pi +\epsilon h(x) \sigma_z
\ea
The exact solutions for the Heisenberg equations of motion for $\phi$ are
\ba
\phi(t,x)= &{1\over 2} \left( \phi(0,x-t)+\phi(0,x+t)
+\int_{x-t}^{x+t}\pi(0,y)dy\right)
 \\
 &-{\epsilon\over 2} \int
\left[\sigma_z(t-|x-y|)-\sigma(-|x-y-t|)\Theta(x-y-t)\right.
 \ne\\
  &~~~~~~~~~~~ \left. -\sigma_z(-|x-y+t|)\Theta(-(x-y+t))\right]
h(y)dy
 \ne
\ea
where $\Theta(x)=\{0 ~if~ x<0; ~1~if~ x>0\}$.

Since in the model,  $\sigma_z$  is a constant of the motion (recall
that I am not taking into account the effects of the operation of the
computer)
 the solution for $\phi$ is thus
\ba
\phi(t,x) =& {1\over 2} \left( \phi(0,x-t)+\phi(0,x+t)
+\int_{x-t}^{x+t}\pi(0,y)dy\right)
\\
 &-{\epsilon\over 2}\sigma_z
\int\left[1-\Theta(x-y-t)-\Theta(-(x-y+t))\right]h(y)dy
h(y) dy
\ea
I will however be working in the Schroedinger representation in the
following.

L assume that  the initial state of the environment is
a thermal density matrix $R_T$ with temperature $T$,
 and the initial state of the spin is a density matrix $\rho(0)$. The
total state is assumed to  be a product state of these two initial states.
 The reduced state
 of the spin system at any time (t) after tracing out over the state of
the
environment is a density matrix
given by
\be
\rho(t)= {1\over 2} (1+\vec\rho(t)\cdot\vec\sigma)
\ee
where $\vec\rho(t)$ is a time dependent vector of length less that or equal
to unity. $\vec \rho(t)$ is given by
\be
\vec\rho(t)= Tr\left(\vec\sigma  e^{i Ht}  \rho(0) R_T e^{-iHt}
\right)
\ee
where the trace is over all of the degrees of freedom of spin system and
bath.

We can write $H$ as
\be
H= e^{i\int \epsilon h(x)\phi(0,x)dx\sigma_z} H_0 e^{-i\int \epsilon
h(x)\phi(0,x)dx\sigma_z}
\ee
since $e^{i\int h(x)\phi(0,x)dx}$ is just the translation operator
taking
$\pi(0,x)$to $\pi(0,x)+\epsilon\int h(x)\phi(0,x)dx\sigma_z$, and since
$\sigma_z$ commutes with $H_0$.

Thus
\ba
\vec\rho(t)=&Tr\left( \vec\sigma e^{i\int \phi(0) \epsilon h dx \sigma_z}e^{-i
\int\tilde \phi(t)\epsilon  h dx \sigma_z}\vec\rho(0)\cdot\vec\sigma \right.
\\ \ne
&~~~~~~~~~~~~~~~\left.\times e^{iH_0 t}R_Te^{-iH_0 t}e^{i\int
\tilde\phi(t)\epsilon
h dx \sigma_z}
e^{-i \int\tilde \phi(0)\epsilon  h dx \sigma_z}\right)
\ea
where $\tilde\phi(t)=e^{iH_0 t}\phi(0,x)e^{-iH_0 t}$ is the time development
of the {\bf free} field with
 the same initial conditions  $\phi(0)$ and $\pi(0)$,
i.e.,
\be
\tilde\phi(t,x)=\half(\phi(0,x-t)+\phi(0,x+t)+\int_{x-t}^{x+t} \pi(0,x')dx')
\ee

Using $\sigma_z^2=1$ and the fact that $R_T$ is diagonal in the energy
representation, we can write $\vec\rho(t)$ as
\be
\vec\rho(t)=Tr\left( \vec\sigma e^{i\int (\phi(0)-\phi(t))\epsilon  h dx
\sigma_z}\vec\rho(0)\cdot\vec\sigma R_Te^{i\int (\tilde\phi(t)-\phi(0))\epsilon
h dx \sigma_z}\right)
\ee
(Note that the extra terms from the Cambell--Baker--Hausdorf formula cancel
out.)
This can furthermore be written as
\ba
\vec\rho(t) =& Tr\left(\vec\sigma (\vec\rho(0)\cdot( \vec\sigma
-(1-cos(\int(\tilde\phi(t)-\phi(0))
\epsilon h dx))\sigma_z \vec e_z \right.
\\ \ne
&~~~~~\left.+ sin(\int(\tilde\phi(t)-\phi(0))\epsilon  h dx)) \vec e_z
\times \vec\sigma R_T\right)
\ea
where $\vec e_z$ is the unit vector in the $z$ direction.
Because $R_T$ is symmetric in $\phi$ and $\pi$, the sin term is zero, and
\be
J(t)\equiv Tr(R_T cos(\int(\tilde\phi(t)-\phi(0))\epsilon  h dx))) = e^{-
\half \int Tr (R_T (\tilde\phi(t)-\phi(0))) \epsilon h dx}
\ee

We are thus left with
\ba
 \rho_z(t)&=& \rho_z(0)
\\
\rho_x(t)&=& J(t) \rho_x(0)
\\
\rho_y(t)&=&J(t) \rho_y(0)
\ea

For later use, let us examine $J(t)$ in various regimes. Let us take $h(x)$
such that
 $h(k)$, the Fourier transform of $h(x)$ is of the form $e^{-\half\Gamma k}$.
 $\Gamma $ is a cutoff
parameter typical of interactions with the environment. I will assume that
$\Gamma>> 1/T$. We then get
\be
ln(J(t))= -{\epsilon^2\over 2} \int \left( {1\over  \pi k} coth({k\over
2T}) (1-cos(kt)) e^{-\Gamma k} \right) dk
\ee
We can approximate $coth(x)\approx 1+e^{-x}({1\over x} +...)$. This gives
us
\be
ln(J(t)) \approx -{\epsilon^2\over 2 \pi} \left(\half ln\left(
{\Gamma^2+t^2\over
\Gamma^2}\right) - \half ln\left( {1+(2Tt)^2}\right) -iTtln\left({ 1-i2Tt\over
1+i2Tt}\right)\right)
\ee

There are essentially three regimes for the time dependence of $J(t)$ given
by the conditions $t<\Gamma$, $\Gamma<t<1/T$ and $t>1/T$.
In the first regime, $t<\Gamma$, we have approximately
\be
ln(J(t)\approx {\epsilon^2t^2\over 4\pi\Gamma^2}
\ee
For the intermediate regime, $\Gamma<t<1/T$ ,the quantum regime, we have
\be
\ln(J(t))\approx -{\epsilon^2\over 2} ln\left(t\over \Gamma\right)
\ee
Finally, for the long time regime $t>>1/T$, the thermal regime, we have
\be
\ln(J(t)) \approx  -\epsilon^2 Tt
\ee
The important feature of these asymptotic formula is that for
the intermediate regime, which I call the quantum regime since the behaviour
is
dominated by the vacuum state of the environment,
$\ln(J)$ increases only logarithmically with $t$. In contrast, the third
regime, the thermal regime, it increases linearly with $t$. This
 will be important in determining the ultimate size of a number which
can be say factored with a quantum computer, because of the dependence
 of the computing time on the length of the number being factored.

This was for the most familiar case of an "ohmic" coupling to the heat
bath. In
the case of superohmic ($h(k(\omega))=\omega^s e^(-\Gamma \omega)$ for
$s>0$), the function $ln(J(t))$ is essentially constant for times less
than $1/T$ and grows
as $t^{1-s}$ in the thermal regime for $s<1$. For $s>1$, $J$ is constant
in both regimes, although it is smaller in the thermal regime than in the
quantum regime. ( and is essentially constant even for such times if $s>1$)
In the subohmic case, $-1<s<0$, on the other hand, $ln J(t)$ grows
roughly as $t^{-s}$ in the quantum regime and as $t^{1-s}$ in the thermal
regime. Again, in the thermal regime the growth in decoherence is a factor
of $t$ larger than in the quantum regime.

The above analysis was carried out for a single bit in the memory
of the quantum computer.
Let us examine the situation in which our memory has some large number
$L$ of bits. Each bit is assumed to couple to its own heat bath of exactly
the above type. The question now is ``What is  the rate of of loss of
 coherence of a coherent sum of
numbers stored in the memory". Ie, define the state $|n> =
|n_{L-1}>|n_{L-2}>...|n_0>$,
where $n_i$ is the ith bit of $n$.
 Consider a coherent state
\be
|\psi>= \sum_n\alpha_n |n>
\ee
The probability that after time $t$ the memory remains in the the state
$\psi$
is given by
\ba
Prob_\psi&=&<\psi|Tr_{environment}
\left(e^{i\sum_iH_it}|\psi>|0><0|<\psi|e^{i\sum_iH_it}\right)
\\
 \ne
&=&\sum_{nn'mm'} \alpha_n^*\alpha_{n'}\alpha_m\alpha_{m'}^*\prod_i
Tr_{environ_i}\left(<n_i|e^{iH_it}|m_i><m'_i|e^{-iH_it}|n'_i>\right)
\\
 \ne
&=& \sum_{nn'}|\alpha_n|^2|\alpha_{n'}|^2\prod_i J_i(t)^{(n_i\otimes n'_i)}
\ea
where $(n_i\otimes n'_i)$ is the XOR of the ith bits of $n$ and $n'$.

This expression tells us how the coherent sum over the various states of
the memory
representing various numbers
decoheres as a function of time. As an example, let us chose the completely
coherent state in which each of the numbers of length $L$ has an equal
probability. This state is typical of the state required in performing
quantum
calculations of the sort in which a quantum computer is much faster than
a classical computer.
Ie, I choose $|\alpha_n|^2= 2^{-L}$. Furthermore let
me assume that each bit is coupled to the environment in exactly the same
way so that $J_i(t)=J(t)$. Then we have
the probability that the coherence will be maintained over time t as

\be
Prob= 2^{-2L}\sum_{nn'} \prod_i J(t)^{(n_i\otimes n'_i)}.
\ee
To evaluate this first fix the number n. The number of numbers n' which
differ from n in
1 bit is $L$. The number which differ in 2 bits is $L(L-1)/2$ and the number
which differ in $r$ bits is ${L!\over r! (L-r)!}$. Thus the above becomes
\be
Prob= 2^{-2L} \sum_n \sum_r {L!\over r! (L-r)!} J^r = 2^{-2L} \sum_n (1+J)^L
= \left( {J+1\over 2}\right) ^L
\ee

If we assume that $1-J$ is very small (which is the only case in which
the system has any hope at all
of acting like a quantum computer), this is well  approximated by
\be
Prob \approx e^{-\half L(1-J)}
\ee
as long as $L(1-J) < 1/(1-J)$.

The strength of the quantum computer is that the time required to perform
the calculation is
a polynomial in the length $L$ of the number.
This time I will designate by $\tau(L)$. Since the quantum calculation
is polynomial in $L$ we can write
$\tau(L)\approx L^a$ for $a>1$. We thus have that the probability of
maintaining coherence over the time of the calculation is of the order
of
\be
\ln(Prob)\approx -O(1)L \epsilon^2 \ln(\tau(N))\approx -O(1) \epsilon^2
L \ln(L)
\ee
in the quantum regime while it is of order
\be
\ln(Prob)\approx -O(1) L^{a+1}\epsilon^2
\ee
with a smooth transition between the two regimes. In order to have a reasonable
probability
of obtaining the correct answer, one needs the probability of obtaining
the quantum coherent
answer to be of order 1. This implies that one must have a sufficiently
small $\epsilon^2$,
the coupling parameter between  the heat bath and
 the system. As long as one is in the quantum regime, the
relation between the coupling $\epsilon^2$ and the
maximum length of the number one can handle
 is essentially inverse linear, no matter what the polynomial
dependence of the calculation.
 However, once one has entered the thermal regime, a decrease in the coupling
buys one only a
small increase in the length of the number $L$ that one can use. I.e.,
in the presence of a coupling to the heat
bath, the thermal time scale ${1\over T}\equiv {\hbar\over k_B T}$ plays
a crucial role. As
long as the calculation can be completed in a time less than this, one
can imagine decreasing
 the coupling to the heat bath for the memory cells so as to achieve the
maximum $L$. If however
the time for the calculation is longer than the thermal time scale, it
becomes very difficult
to decrease the coupling to the bath sufficiently to achieve the necessary
coherence.

Is it possible to use the computer even if the quantum state looses
coherence?  I cannot answer this in general, but can show that one
strategy does not work.
One could  imagine trying to make up for the loss of coherence by increasing
the number
of times the program is run. (This is in fact a crucial factor in the
Shor algorithm for factoring, not because of decoherence, but because the
calculation itself has a finite probability of not giving the correct
outcome.)  After a sufficient number of attempts, one should by chance
have a system which has maintained coherence.
 In the factoring problem, one can test ones answer ( does it give the
factors of the number),
 and simply keep repeating the experiment until one gets the right answer.
However, in M trials,
the probability of never finding a coherent outcome is $(1-Prob)^M\approx e^{-M
{}~Prob}$.
The
number of trials required to make this small (i.e., so that one has a
high probability of having had a coherent run) is thus , the required
number of attempts is $M\approx 1/Prob\approx e^{O(1)Lln(L)} $
in the quantum regime, which is exponential in the length. In the thermal
regime, this time
scale is even worse. One will thus have lost all advantages of the
quantum nature of the computer.
We see that one must  make sure that coherence is maintained during the
calculation.

In order to maintain coherence, one must have a small value for
$\epsilon^2$. At first as one decreases $\epsilon^2$, the gain in the
maximum length number one can factor is roughly inversely proportional
to the value of $\epsilon^2$. However, once $\epsilon$ is sufficiently
small that the time scale of computation for
 the maximum length which can maintain coherence approaches the inverse
thermal time $1/T$, one reaches a bottleneck. Further reductions in
$\epsilon^2$ now have little effect on the maximum length. The
decoherence due to the rapidly increasing time spent in computation
cancels out the effect of the smaller $\epsilon^2$. Thus the thermal
time scale $1/T$ sets an effective limit to the time of the calculation,
and thus a weak limit on the maximum length of the numbers one can
compute with.

 If one imagines factoring a 1000 bit number, and one assumes that the
quantum factoring time
can be made to be of order $L^2$ (probably the slowest rate imaginable),
we find that one must carry out at least $10^6$ calculation in
 the thermal time scale. Since the thermal time scale for a temperature
of 1K is of the order of $10^{-9}$ sec, this
would imply that one would have to use a computer which ran at optical
 frequencies.

\section{Other Noise}

 The above coupling to the heat bath is "error free" in the sense that
if
one is in a number eigenstate (ie, is in a state $|n>$), the system will
remain in that state throughout. The environment does not cause spin
flips.
 What about the situation in which there is also some probability
of a state flip- ie of the system making a transition between the two
eigenstates
of $\sigma_z$? One could approximate this by assuming that the coupling
to the heat bat is via say
$$\sigma_\theta=\cos(\theta) \sigma_z +\sin(\theta)\sigma_x$$, with small
$\theta$.

The above analysis is exactly the same for this case, where we replace
$\sigma_z$ everywhere by $\sigma_\theta$.
Writing the number eigenstates with respect to $\sigma_\theta$ so that
\be
|n>_\theta= |n_{L-1}>_\theta ....|n_0>_\theta
\ee
we have
\be
|n>= \sum_m \prod_i \cos(\theta)^{1\otimes n_i\otimes m_i}
sin(\theta)^{n_i\otimes m_i} (-1)^{(n_i\otimes m_i)n_i} |m>_\theta
\ee
The probability of remaining in the state $|n>$ under the coupling to the
heat bath is then
\be
Prob= \sum_m\sum_m' \cos(\theta)^{2S(n,m)}(J\sin(\theta)^2){(L-S(n,m))}
\ee
where $S(n,m)$ is the number of bits in which $n$ and $m$ are the same.
Again using the arguments above as to the number of terms where the $S$
has a given value, we get
$$Prob= (cos(\theta)^2 + J sin(\theta)^2)^L$$
For small $\theta$ this gives
\be
Prob= (1 - (1-J) \theta^2)^L \approx e^{-L\theta^2 J}.
\ee
Thus $\theta$ must be kept very small in order to ensure that the probability
of error remains small. However we note that the probability of error is
vastly suppressed with respect to the decoherence probability, which is
in accord with the observation that the decoherence effects are in general
much larger and more rapid than are transition effects.

This has assumed that the process causing spin flips is the same as the
one causing loss of coherence in a superposition of the two spin states.
  In general, the environmental degrees
of freedom which cause decoherence are not the same as those causing bit
flips.
I will therefore look at the alternative situation in which the single
bit
Hamiltonian is of the form
\be
\half \left( (\pi_1-\epsilon_1 h(x)\sigma_z))^2 +(\partial_x \phi_1)^2
+
(\pi_2-\epsilon_2 h(x)\sigma_x))^2 +(\partial_x \phi_2)^2 (\pi_3-\epsilon_2
h(x)\sigma_y))^2 +(\partial_x \phi_3)^2\right)
\ee

Since we want the single bit decoherence and bit flip probabilities to
be small ( or else the quantum computer is useless from the  start), I
will assume that
the $\epsilon_k$ are all sufficiently small. Furthermore, for simplicity
 I will
take  $\epsilon_2=\epsilon_3$, so that the spin flip processes are of
equal strength. I cannot solve this problem exactly, but since the
probabilities are assumed to be very small,
 one can calculate the transition probability to lowest order
in the various epsilons. The Hamiltonian can be written as
\be
H=H_0 - \sum_i \left(\int\epsilon_i \pi_i(x) \sigma_i h(x)dx+ \half \int
h(x)^2 dx\right)
\ee
where $\sigma_i=(\sigma_z,\sigma_x,\sigma_y)$.
The reduced density vector $\vec\rho(t)$ is given by
\be
\vec\rho(t) = Tr\left(\vec\sigma  e^{iHt}\vec\rho(0)\cdot\sigma R_T e^{-iHt}
\right)
\ee

To zeroth order, since $H=H_0$  is independent of $\sigma$, we have
$\rho(t)=\rho(0)$.
To first order, one obtains terms which are linear in the $\pi$s and the
$\phi$s. However in the thermal state, all of these are zero, because the
thermal state (of $H_0$) is symmetric in the fields. To second order the
results are non-zero. However all of the cross terms $\epsilon_i\epsilon_j$
for $i\neq j$ will again be zero because the fields are by assumption
independent
and thus the cross correlations between terms linear in each of the fields
will again be zero. Thus the only terms surviving will be the terms
proportional
to  $\epsilon_i^2$. But each of  these terms are independent of the other
$\epsilon$s.
I.e., each of these terms are the same as those obtained by setting the
other two epsilons to zero. These are however just the same as the second
order terms calculated above in the first part. We thus get
\ba
\rho(t)_i =& \sum_j\left(\delta_{ij}
\left(1-\half \sum_k \left(\epsilon_k^2Tr(R_T (\int(\phi_k(t)-\phi_k(0))hdx)^2
)\right)\right) \right.
\\
 \ne
&+\half\sum_k\epsilon_k^2 Tr(R_T (\int(\phi_k(t)-\phi_k(0))hdx)^2 )
\delta_{ik}\delta_{jk}
\left.\right)\rho_j(0)
\ea
Note that since all of the fields are  of the same form and at the same
temperature, the $Tr(R_T(\int \phi_i(t)-\phi_i(0) )h dx)^2)$ are the
same for all $i$.

The probability of bit flip then becomes

\be
Prob_{noflip}\approx (1-\epsilon_2^2 <0| (\phi(t)-\phi(0))^2 |0>)^L
\ee
while the probability of decoherence for a state which is the coherent
sum over all the integers of length L is  given by
\be
Prob_{decoher}\approx (1-(\half(\epsilon_1^2+\epsilon_2^2)
<0|(\phi(t)-\phi(0))^2|0>
\ee
If $\epsilon_1>>\epsilon_2$, the decoherence will again be much more rapid
that the probability of "error" due to bit flip.

\acknowledgments
I would especially like to thank the Santa Fe Institute and the organizers
 of the May 1994 conference on Complexity, Entropy, and the Physics of
Information where the issue of quantum computing was a key theme, and
which incited my interest in the problems thereof.

\section{Conclusion}

Quantum computation places the demand on the system that the coherence
of the initial state be maintained throughout the computation. In order
to maintain
 this coherence in the presence of a heat bath, the reduction in the coupling
to
the heat bath buys one a proportional increase in the size of the computation
only
in the computation can be completed within a thermal time scale. For
computation
times longer than the thermal time scale, a decrease in the coupling gives
one relatively
little change in the size of the  possible coherent computation.
The thermal time scale thus sets a (weak) limit on the length of time
that a quantum calculation can take.

\end{document}